# Alternative scenarios for the LHC based electron-proton collider

A. N. Akay[a]  B. Dagli[a]  B. Ketenoglu[b,*]  A. Ozturk[a,c]  S. Sultansoy[a,d]

[a]TOBB University of Economics and Technology, Ankara, Turkey
[b]Department of Engineering Physics, Ankara University, Ankara, Turkey
[c]Turkish Accelerator and Radiation Laboratory, Ankara, Turkey
[d]ANAS Institute of Physics, Baku, Azerbaijan

**Abstract:** Construction of the ERLC (twin LC) collider tangential to LHC will give opportunity to investigate *ep* collisions at essentially higher center-of-mass energies than ERL50 and LHC based *ep* collider. Luminosity estimations show that values well exceeding $10^{34}$ cm$^{-2}$s$^{-1}$ can be achieved for ERLC and HL-LHC based *ep* colliders. Certainly, proposed *ep* colliders have great potential for clarifying QCD basics and new physics search in addition to providing precise PDFs for adequate interpretation of the LHC experimental data. Another alternative to the ERL50 is to construct an *e*-ring with the same energy and length. In this case, luminosity of order $10^{34}$ cm$^{-2}$s$^{-1}$ can be reached. The advantage of this option is that the *μ*-ring can be installed instead of the *e*-ring as a next stage, which will allow it to reach a much higher center of mass energy.

## 1. INTRODUCTION

It is known that lepton-hadron collisions played crucial role in our understanding of deep structure of matter: proton form-factors were first observed in electron scattering experiments, quark-parton model was established at SLAC deep inelastic electron scattering experiments and so on. The first and still unique electron–proton collider HERA, further explored structure of protons and provided parton distribution functions (PDFs) for Tevatron and the LHC.

Let us emphasize that the Higgs mechanism is responsible for less than 2% of the mass of the visible universe, remaining 98% is provided by strong interactions. Exploration of region of $x \ll 10^{-4}$ at $Q^2 \gg 10$ GeV$^2$ is very important for clarification of QCD basics: first condition corresponds to huge parton densities (gluon saturation region), the second condition ensures the perturbation regime. Therefore, lepton-hadron colliders with $\sqrt{s} > 1$ TeV are strongly needed.

Unfortunately, HERA ($\sqrt{s_{ep}} \approx 0.3$ TeV) did not reach this region, *e*-RHIC with $\sqrt{s_{ep}} \approx 0.1$ TeV is not effective for this region even with *eA* option. The HERA *eA* option could have touched this region, but unfortunately, it was not implemented.

It is known that construction of future linear colliders (or dedicated *e*-linac) tangential to energy frontier hadron colliders will give opportunity to investigate lepton-hadron interactions at highest center-of-mass energies (see review [1] and references therein). This idea was first proposed [2,3] in mid-1980s for combination of UNK *pp* and VLEPP *e*$^+$*e*$^-$ colliders. Then, THERA (TESLA on HERA) proposal [4] was developed in the end of 1990s. Concerning *e*-linac and LHC based *ep* colliders, the most advanced proposal was the ILC-LHC based QCD-Explorer [5], which has later turned into LHeC [6]. This idea was applied for FCC and SppC in [7] and [8], respectively. Nowadays, muon-hadron colliders have come into agenda (see review [9] and references therein) with the possibility of construction of muon collider (or dedicated muon ring) tangential to existing and proposed high energy hadron colliders.

---

*e-mail: bketen@eng.ankara.edu.tr

In Table 1, we present a correlation between colliding beams and collider schemes in energy-frontier aspect.

Table 1. Energy frontier colliders: colliding beams vs. collider types.

| Colliders | Ring-Ring | Linac-Linac | Linac-Ring |
|---|---|---|---|
| Hadron | + | | |
| Lepton ($e^-e^+$) | | + | |
| Lepton ($\mu^-\mu^+$) | + | | |
| Lepton-hadron ($eh$) | | | + |
| Lepton-hadron ($\mu h$) | + | | |
| Photon-hadron | | | + |

Among the proposed $eh$ colliders, $e$-RHIC [10] (approved by DoE) and LHeC at CERN are scheduled to be constructed in 2030s. Let us mention that $ep$ center-of-mass energy at $e$-RHIC ($\sqrt{s_{ep}} = 105$ GeV) is three times lower than HERA (main advantages of the eRHIC are $eA$ collisions and polarized beams). The version of LHeC approved in 2010s is based on 60 GeV energy recovery electron linac (ERL60) and corresponds to $\sqrt{s_{ep}} = 1.3$ TeV which is four times higher than that of HERA. *It should be emphasized that ERL version does not provide the opportunity to increase ep center-of-mass energy.*

Since the LHC provides the highest energy proton beams, construction of $e$-linac or $e/\mu$-ring tangential to the LHC will provide highest CoM energies in lepton-hadron collisions. Let us mention that the first LHC based $ep$ collider proposal was LEP⊗LHC [11]. However, this option was excluded due to problems with synchronous operation of LEP and LHC. LHC based linac-ring type $ep$ colliders were proposed 30 years ago [12-14]. This option came into agenda again in the beginning of 2000s [15]. Later on, international cooperation on LHC-based $ep$ colliders was established and the results of the studies were published in 2012 [6], which includes 3 different options: 140 GeV single-pass linac, ERL60 and new $e$-ring in the LHC tunnel. In our opinion, work on the last option was senseless for the same reasons that giving up LEP⊗LHC. On the other hand, even though ERL60 provides higher luminosities, this option does not give the opportunity to increase of energy as mentioned above. In further phase, energy of ERL was diminished to 50 GeV from 60 GeV [16]. Finally, it was proposed to construct a new $e$-ring tangential to the LHC with the same length and energy as ERL60, giving the same $ep$ luminosity [17].

As mentioned above, exploration of region of $x \ll 10^{-4}$ at $Q^2 \gg 10$ GeV$^2$ is very important for clarification of QCD basics. LHeC with ERL50 ($\sqrt{s_{ep}} = 1.18$ TeV) will give opportunity to start exploration of this region (see Table 7). Certainly, higher CoM energies will be useful for search of QCD basics in depth. For linac-ring option, this means turning into single-pass linac choice. Concerning ring-ring option, 50 GeV $e$-ring may further be upgraded to 1.5 TeV muon ring.

Recently, V. I. Telnov has proposed ERLC (twin LC) scheme to improve ILC luminosity by two orders [18]. In our opinion, Telnov's proposal will have an essential effect upon decision on construction of the ILC since the luminosity of ERLC is comparable with that of the FCC-*ee*.

Certainly, Telnov's proposal is important for linac-ring type electron-hadron colliders as well. For this reason, ERLC and LHC based *ep* and *eA* colliders have been considered in [19] and [20], respectively. However, 125 GeV electron beam parameters for ERLC available at that time have been used (see the first version of the arXiv preprint [21]). In the last version published in Journal of Instrumentation, 125 GeV electron beam parameters have been modified and 250 GeV option has been added [18].

In this paper, we have re-estimated parameters of the ERLC and LHC based *ep* colliders as well as a new 50 GeV *e*-ring based one. Section 2 is devoted to brief description of the AloHEP software, which is used for evaluation of *ep* collider parameters. Main parameters of ERLC and LHC that are used in luminosity and beam-beam tune shift calculations are presented in Section 3. Evaluation of parameters of ERLC and LHC based *ep* colliders is given in Section 4. Main parameters of 50 GeV *e*-ring and LHC based *ep* collider are discussed in Section 5, where we also present parameters of *μp* collider that may be considered as a next stage. Section 6 contains several remarks on physics search potential of the LHC-based lepton-proton colliders. Finally, our conclusion and recommendations are given in Section 7.

## 2. THE ALOHEP SOFTWARE FOR LUMINOSITY ESTIMATIONS

Center-of-mass energies and luminosities of colliders are the most important parameters in terms of physics search potential. Corresponding equations are given below:

$$\sqrt{s} = 2\sqrt{E_1 E_2}$$

$$\mathcal{L} = \frac{N_1 N_2}{4\pi \, max[\sigma_{x_1}, \sigma_{x_2}] \, max[\sigma_{y_1}, \sigma_{y_2}]} \, min\,[f_{c_1}, f_{c_2}]$$

where $E_1$ and $E_2$ are the colliding beam energies, $N_1$ and $N_2$ are the number of particles in corresponding bunches, $\sigma_{x,y}$ are transverse beam sizes, $f_{c_1}$ and $f_{c_2}$ are the bunch frequencies at interaction point.

In addition, beam-beam interactions have to be kept under control: in case of linac-ring type eh colliders they are disruption for linac electrons and tune-shift for ring hadrons. Corresponding equations are given below:

$$D_{x_e} = \frac{2Z_h N_h r_e \sigma_{z_h}}{\gamma_e \sigma_{x_h}(\sigma_{x_h} + \sigma_{y_h})}$$

$$D_{y_e} = \frac{2Z_h N_h r_e \sigma_{z_h}}{\gamma_e \sigma_{y_h}(\sigma_{x_h} + \sigma_{y_h})}$$

$$\xi_{x_h} = \frac{N_e r_h \beta^*_{x_h}}{2\pi \gamma_h \sigma_{x_e}(\sigma_{x_e} + \sigma_{y_e})}$$

$$\xi_{y_h} = \frac{N_e r_h \beta^*_{y_h}}{2\pi \gamma_h \sigma_{y_e}(\sigma_{y_e} + \sigma_{x_e})}$$

where $N_e$ and $N_h$ are particle numbers in electron and hadron bunches, respectively; $Z_h$ is 1 for proton; $r_e$ and $r_h$ are classical radius of electron and hadron, respectively; $\sigma_{z_h}$ is length of hadron bunch; $\beta^*_{x_h}$ and $\beta^*_{y_h}$ are horizontal and vertical beta-functions at interaction point; $\gamma_e$ and $\gamma_h$ are Lorentz factors of electron and hadron, respectively.

These parameters can be analytically calculated by neglecting some special cases that can cause changes in the shape of the beam in the collision area. To obtain more accurate results by taking into account these effects, we need to perform the calculation numerically with the help of a simulation. With this purpose the software AloHEP has been developed several years ago for estimation of main parameters of linac-ring type *ep* colliders [22, 23]. Recently, AloHEP has been upgraded [24] for all types of colliders (linear, circular and linac-ring) as well as colliding beams (electron, positron, muon, proton and nuclei).

## 3. PARAMETERS OF ERLC AND LHC

In this section, we present parameters of LHC and ERLC, which are used for estimation of main parameters of *ep* colliders in the following sections 4 and 5. Parameters of HL-LHC proton beam upgraded for ERL-50 related *ep* collider [16] are given in Table 2 (see Tables 3, 6 and 35 in [16] Run 6 option).

Table 2. HL-LHC proton beam parameters upgraded for ERL-50 based *ep* collider

| Parameter [unit] | HL-LHC |
|---|---|
| Beam Energy [TeV] | 7 |
| # of protons per bunch [$10^{11}$] | 2.2 |
| Norm. emit., $\epsilon_n$ [μm] | 2.5 |
| $\beta_{x,y}$ at IP [m] | 0.07 |
| $\sigma_{x,y}$ at IP [μm] | 4.85 |
| Bunches per beam | 2748 |
| Bunch spacing [ns] | 25 |
| Bunch length [mm] | 90 |

Table 3.1. Main parameters of electron beam of $\sqrt{S}$=250 GeV ERLC and ILC

| Beam Energy [GeV] | 125 | |
|---|---|---|
| Collider | ERLC | ILC |
| # of electrons per bunch [$10^9$] | 1.77 | 20 |
| Norm. emit., $\epsilon_x$ [μm] | 10 | 5 |
| Norm. emit., $\epsilon_y$ [μm] | 0.035 | 0.035 |
| $\beta_x$ at IP [cm] | 6.8 | 1.3 |
| $\beta_y$ at IP [cm] | 0.031 | 0.04 |
| $\sigma_x$ at IP [μm] | 1.66 | 0.52 |
| $\sigma_y$ at IP [nm] | 6.2 | 7.7 |
| Rep. rate [Hz] | 6.5x$10^8$ | 6560 |
| Bunch distance [m] | 0.46 → 0.5* | 166 →165* |
| $\sigma_z$ at IP [cm] | 0.03 | 0.03 |

Parameters of ERLC and ILC are given in Tables 3.1 and 3.2 (see Tables 2 and 3 of Reference [18]. For electron-hadron colliders we have used ERLC's *cw* option with 0.65 GHz RF frequency).

Table 3.2. Main parameters of electron beam of $\sqrt{S}$=500 GeV ERLC and ILC

| Beam Energy [GeV] | 250 | |
|---|---|---|
| Collider | ERLC | ILC |
| # of electrons per bunch [$10^9$] | 1.23 | 20 |
| Norm. emit., $\epsilon_x$ [μm] | 10 | 10 |
| Norm. emit., $\epsilon_y$ [μm] | 0.035 | 0.035 |
| $\beta_x$ at IP [cm] | 9.4 | 1.1 |
| $\beta_y$ at IP [cm] | 0.089 | 0.04 |
| $\sigma_x$ at IP [μm] | 1.38 | 0.47 |
| $\sigma_y$ at IP [nm] | 7.4 | 5.9 |
| Rep. rate [Hz] | 6.5x$10^8$ | 6560 |
| Bunch distance [m] | 0.46 → 0.5* | 166 →165* |
| $\sigma_z$ at IP [cm] | 0.089 | 0.03 |

*We use these values to synchronize electron bunch distances with *p* bunch distances.

## 4. ERLC AND HL-LHC BASED *ep* COLLIDERS

In this section, we evaluate main parameters of the ERLC and HL-LHC based *ep* colliders, namely, luminosity, disruption and beam-beam tune-shift by using the AloHEP software and nominal parameters of proton and electron beams given in Tables 2 and 3, respectively. As a result, we have obtained the parameters of the corresponding *ep* colliders given in Table 4. It is seen that ERLC option results in more than 2 orders of higher luminosity values compared to ILC.

Table 4. Main parameters of the ERLC and HL-LHC based *ep* colliders

| | ERLC | | ILC | |
|---|---|---|---|---|
| $E_e$ [GeV] | 125 | 250 | 125 | 250 |
| $\sqrt{S}$ [TeV] | 1.87 | 2.65 | 1.87 | 2.65 |
| $L$x$10^{30}$ [cm$^{-2}$s$^{-1}$] | 4080 | 2840 | 9.79 | 9.79 |
| Disruption, $D_e$ | 9.72 | 4.86 | 9.72 | 4.86 |
| $\xi_p$ [$10^{-4}$] | 0.86 | 0.60 | 9.8 | 9.7 |

While center-of-mass energies of ERLC based *ep* colliders are higher than that of ERL-50 based ones (see Table 5), luminosity values are several times lower. For this reason, we have also added ERLC* option with upgraded ERLC parameters: 15 times lower repetition rate and 15 times higher number of electrons per bunch, since with nominal ERLC bunch distance (0.5 m) only 1/15 of electron bunches collide with proton bunches (bunch distance 7.5 m). As a result, luminosity of ERLC*-125 and ERLC*-250 based *ep* colliders exceed that of ERL-50 based one by factors 6.8 and 4.7, respectively. In order to compare, we give center-of-mass energy, luminosity, disruption and beam-beam tune-shift values of ERL-50 based *ep* colliders as well [16].

Table 5. Main parameters of the ERL-50, ERLC*-125, ERLC*-250 and HL-LHC based *ep* colliders

|  | ERL | ERLC* | |
|---|---|---|---|
| $E_e$ [GeV] | 50 | 125 | 250 |
| $\sqrt{s}$ [TeV] | 1.18 | 1.87 | 2.65 |
| L [$10^{33}$ cm$^{-2}$s$^{-1}$] | 9 | 61.2 | 42.6 |
| Disruption, $D_e$ | 14.5 | 9.72 | 4.86 |
| Tune shift, $\xi_p$ [$10^{-4}$] | 1.5 | 13.0 | 9.0 |

Let us mention that luminosity of the ERLC based *ep* colliders can be further improved using dynamic focusing scheme [25]. This scheme should be adopted for vertical plane, since vertical size of the ERLC bunch (Table 3) is three orders less than vertical size of HL-LHC bunch (Table 2), while horizontal sizes are of the same order. After all, $L_{ep} \gtrsim 10^{35}$ cm$^{-2}$s$^{-1}$ seems to be achievable with reasonable modifications on ERLC and HL-LHC parameters.

## 5. *e*-RING BASED *ep* COLLIDER

In this section, we evaluate main parameters of the 50 GeV *e*-ring and LHC based *ep* collider. The tunnel constructed for the electron ring can then be used to implement the 1.5 TeV muon ring. Thus, a muon-proton collider with a much higher center of mass energy can be realized as the next stage.

### 5.1. Main parameters of electron-proton collider

Tentative parameters for *e*-ring are presented in Table 6. Circumference of *e*-ring is chosen to be equal to total length of ERL-50. At this stage we consider round electron bunches and require that transverse sizes of electron and proton bunches are matched at interaction point (IP). Assuming that electron beam $\beta$ function at IP is 5 cm, this requirement leads to $\epsilon_N = 46$ μm in order to obtain 4.85 μm transverse beam size (see Table 2). Number of electrons per bunch and number of bunches in *e*-ring corresponds to 100 mA beam current (Table 6.1 in [6]).

Table 6. Electron beam parameters for *e*-ring

| Beam Energy [GeV] | 50 |
|---|---|
| Circumference [km] | 5.4 |
| Number of electrons per bunch [$10^{10}$] | 2.0 |
| Number of bunches | 562 |
| Norm. emit., $\epsilon_n$ [μm] | 46 |
| $\beta$ at IP [cm] | 5.0 |
| $\sigma$ at IP [μm] | 4.85 |
| Bunch distance [m] | 7.5 |

Implementing parameters from Tables 2 and 6 into AloHEP software, we obtained parameters of *ep* collisions given in second column of Table 7. It is seen that electron beam-beam tune shift value is unacceptably high.

Table 7. Main parameters of the *e*-ring and HL-LHC based *ep* collider

| Parameter [unit] | 50 GeV *e*-ring | |
|---|---|---|
|  | Nominal | Upgraded |
| $\sqrt{s}$ [TeV] | 1.18 | |
| L [$10^{33}$ cm$^{-2}$s$^{-1}$] | 46 | 4.2 |
| Tune Shift, $\xi_e$ | 1.07 | 0.1 |
| Tune Shift, $\xi_p$ [$10^{-4}$] | 9.8 | 9.8 |

One way to reduce $\xi_e$ down to 0.1 is reducing $N_p$ from $2.2 \times 10^{11}$ to $2.0 \times 10^{10}$ which results in a corresponding decrement of luminosity value. This option is presented in the last column of Table 7. Therefore, realistic value for luminosity is $4.2\times10^{33}$ cm$^{-2}$s$^{-1}$.

### 5.2. Muon-proton collider as a next stage

One can consider the two-stage scenario for the LHC-based lepton-hadron colliders: the LHeC option with 5.4 km $e$-ring as the first stage, already tangential to the LHC, followed by the construction of $\mu$-ring in the same tunnel. The LHC-based muon-proton colliders have been proposed in [26]. The 1.5 TeV muon beam option (see Table 4 in [26]) is most suitable for the proposed scenario. This option will provide 6.48 TeV center-of-mass energy and $2.1\times10^{33}$ cm$^{-2}$s$^{-1}$ luminosity.

Another option based on existing hadron colliders is RHIC-based muon-proton colliders [27, 28] (for a comparison of the main parameters of the muon-hadron colliders proposed so far, see review [9]). As expected, the center-of-mass energy and luminosity of RHIC-based $\mu p$ colliders are an order lower than LHC-based ones (see Table 28 in [9]).

It should be mentioned that earlier construction of muon-hadron colliders is more probable than construction of muon colliders since requirements on muon beam parameters are more tolerable (smaller number of muons and higher emittance).

### 6. BRIEF REMARKS ON PHYSICS SEARCH POTENTIAL

Construction of energy frontier $ep$ colliders is a must to provide precision PDFs for adequate interpretation of the HL-LHC and FCC/SppC experimental data. Another goal, in our opinion even more important, is clarifying the QCD basics, especially understanding of confinement. In this context two regions of $x$ Bjorken are crucial: small $x$ at high $Q^2$ and $x \approx 1$. Highest energy is important for the former and highest luminosity for the latter. Achievable $x$ values at $Q^2 = 100$ GeV$^2$ and $Q^2$ values at $x=10^{-5}$ are presented in Table 8. Obviously, the ERLC and $\mu$-ring options will allow detailed examination of the gluon saturation region, while the ERL50 and $e$-ring options will cover the starting zone of this region only.

Table 8. Achievable $x$ Bjorken values at $Q^2=100$ GeV$^2$ and $Q^2$ values at $x=10^{-5}$.

|  | $x$ at $Q^2 = 100$ GeV$^2$ | $Q^2$ at $x=10^{-5}$ |
|---|---|---|
| ERL50 | $7.2\times10^{-5}$ | 14 GeV$^2$ |
| ERLC-125 | $2.9\times10^{-5}$ | 35 GeV$^2$ |
| ERLC-250 | $1.4\times10^{-5}$ | 70 GeV$^2$ |
| $e$-ring 50 | $7.2\times10^{-5}$ | 14 GeV$^2$ |
| $\mu$-ring 1500 | $2.4\times10^{-6}$ | 420 GeV$^2$ |

High luminosity is crucial for investigation of the Higgs boson properties at $ep$ colliders as well. Obviously, usage of ERLC* instead of ERL50 is advantageous since it provides higher center of mass energies and luminosities. Concerning BSM physics, proposed colliders have great potential for a lot of topics such as: leptoquarks, excited electron and neutrino, color octet electron, contact interactions, SUSY, RPV SUSY (especially resonant production of squarks), extended gauge symmetry (especially left-right symmetric models) etc.

The importance of the Higgs boson for the current state of the universe is that it provides the masses of the $u$, $d$ quarks and the electron. For example, if the $d$ quark were lighter than the $u$ quark or if the difference between the mass of $d$ and $u$ were smaller than the mass of the electron, neither we could have written this article, nor you would have been able to read it. Unfortunately, even with super high luminosity, LHeC will not allow investigation of Higgs interactions with the first SM family fermions. *Therefore, LHeC is a QCD Explorer, not a Higgs Factory.*

## 7. CONCLUSION

It is shown that construction of ERLC (twin LC) tangential to LHC will give opportunity to realize multi-TeV center-of-mass energy *ep* colliders with luminosities around $10^{35}$ cm$^{-2}$s$^{-1}$. Certainly, these colliders will essentially enlarge physics search potential of the LHC for both the SM and BSM phenomena. In particular, clarification of QCD basics, including confinement, is one of the most important goals.

The resolution power of HERA, *e*-RHIC, ERL-50 based LHeC, ERLC⊗LHC *ep* colliders as well as *µ*-ring and LHC based *µp* collider is presented in Figure 1 (SPL stands for 125 GeV Single Pass Linac; ERLC125 and ERLC250 correspond to ERLC with beam energies of 125 GeV and 250 GeV, respectively; *µ*-ring corresponds to 1500 GeV muon beam). The advantage of ERLC and *µ*-ring over ERL-50 is obvious. Let us remind that ERLC* provide essentially higher luminosity values as well (see Table 5). In addition, ERLC and *µ*-ring will allow us to better examine the small *x* Björken region (see Table 8).

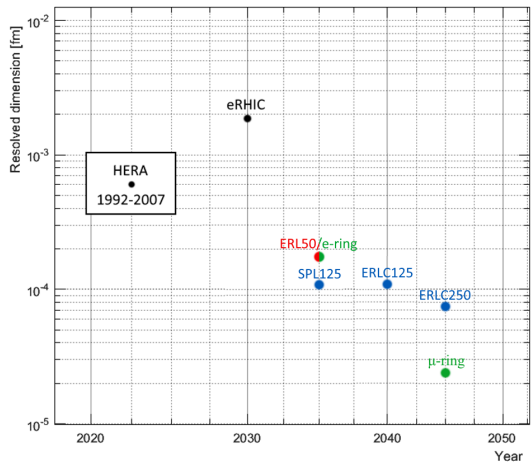

Figure 1. The resolution power of the LHC-based lepton-proton colliders.

Recently, parameters of the ERL and the LHC based *ep* collider has been upgraded once more [29]. It is mentioned above that electron beam energy was diminished from 60 GeV [6] to 50 GeV [16]. Considering the latest upgrade, circumference is increased from 5.4 km to 6.7 km together with moderate modifications of electron and proton beam parameters, which results in a decrement of luminosity down to $3 \times 10^{33}$ cm$^{-2}$s$^{-1}$ [29]. Corresponding modification of 50 GeV *e*-ring parameters will provide the opportunity to achieve luminosity values exceeding $10^{34}$ cm$^{-2}$s$^{-1}$ [30] for this option of the LHeC.

In conclusion, it is obvious that the construction of the LHC based *ep* colliders is a must for a number of reasons. Let us remind that ERL-50 option does not provide the opportunity to increase center-of-mass energy. For this reason we propose two possible alternative scenarios:

- Concerning linac-ring type *ep* colliders, following scenario may be realized: construction of 125 GeV single pass *e*-linac tangential to LHC as a first stage, followed by ERLC with √s = 250 GeV as a second stage to enhance luminosity, with a further upgrade to ERLC with √s = 500 GeV (or higher if necessary).
- Second alternative is construction of new 50 GeV *e*-ring tangential to LHC following by construction of 1.5 TeV *µ*-ring in the same tunnel. Physics search potential of the *e*-ring option almost coincides with the ERL-50 option. Certainly, *µp* collider with much higher center-of-mass energy will provide huge search potential. Since *µp* collider will be the next stage in the *e*-ring part, it should be installed in the deepest and least populated area of the LHC.

Finally, we believe that systematic studies of accelerator, detector and physics search aspects of ERLC (as well as 50 GeV *e*-ring) and LHC based electron-hadron colliders are essential for long-term planning of High Energy Physics.

**Acknowledgements**

The authors are grateful to D. Akturk and U. Kaya for useful discussions.

**References**

[1] Akay, A. N., Karadeniz, H., & Sultansoy, S., *International Journal of Modern Physics A*, **25(24)**, 4589 (2010).

[2] Alekhin, S. I., Boos, E., Borodulin, V. I., et al., Preprint IFVE-87-48 (1987).

[3] Alekhin, S. I., Boos, E. E., Borodulin, V. I., et al., *International Journal of Modern Physics A*, **6**, 21 (1991).

[4] H. Abramovitz et al.: in TESLA TDR, DESY-2001-011, ECFA-2001-209 (2001).

[5] Sultansoy, S. *The European Physical Journal C-Particles and Fields*, **33**, s1064 (2004).

[6] Fernandez, J. A., Adolphsen, C., Akay, A. N., et al. *Journal of Physics G: Nuclear and Particle Physics*, **39(7)**, 075001 (2012).

[7] Acar, Y. C., Akay, A. N., Beser, S., et al. *Nuclear Instruments and Methods in Physics Research Section A: Accelerators, Spectrometers, Detectors and Associated Equipment*, **871**, 47 (2017).

[8] Canbay, A. C., Kaya, U., Ketenoglu, B., et al. *Advances in High Energy Physics*, **2017** 4021493 (2017).

[9] Ketenoğlu, B., Dağlı, B., Öztürk, A. and Sultansoy S., *Modern Physics Letters A*, **37(37n38)**, 2230013 (2022).

[10] Khalek, R. A., Accardi, A., Adam, J., et al. *Nuclear Physics A*, **1026**, 122447 (2022).

[11] Brianti, G. No. CERN-LHC-Note-134. CM-P00063543 (1990).

[12] Tigner, M., Wiik, B. and Willeke F. Proceedings of IEEE 1991 Particle Accelerator Conference, San Francisco, California, pp. 2910-2912 (1991).

[13] Grosse-Wiesmann, P. *Nucl. Instrum. Meth. A* **274**, 21 (1989).

[14] Sultanov, S., ICTP Preprint IC/89/409 Trieste (1989).

[15] Sultansoy, S. Presentation at Miniworkshop on Machine And Physics Aspects of CLIC Based Future Collider Options, http://cds.cern.ch/record/798726/files/open-2004-026.pdf (2004)

[16] Agostini, P., Aksakal, H., Alekhin, S., et al. *Journal of Physics G: Nuclear and Particle Physics*, **48(11)**, 110501 (2021).

[17] Kaya, U., Ketenoğlu, B., Sultansoy, S. *Süleyman Demirel University Faculty of Arts and Science Journal of Science* **13(2)** (2018); e-Print: 1710.11157 [physics.acc-ph].

[18] Telnov, V. I. *Journal of Instrumentation*, **16(12)**, P12025 (2021).

[19] Dagli, B., Ketenoglu, B., & Sultansoy, S. *ArXiv Preprint:2107.04850* (2021).

[20] Akay, A. N., Dagli, B., Ketenoglu, B., et al. *ArXiv Preprint:2107.08312* (2021).

[21] Telnov, V. I. *Journal of Instrumentation*, **16**(12), P12025 (2021).

[22] yefetu. *ALOHEP sofware* [Java]. https://github.com/yefetu/ALOHEP (2022).

[23] yefetu. *AloHEP Sofware*. http://yef.etu.edu.tr/ALOHEP_eng.html (2021)

[24] Dagli, B., Sultansoy, S., Ketenoglu, B., & Oner, B. B. In *12th Int. Particle Accelerator Conf.* (2021).

[25] Brinkmann, R., & Dohlus, M. (No. DESY-M-95-11) SCAN-9511094. (1995).

[26] Kaya, U., Ketenoglu, B., Sultansoy, S. and Zimmermann, F. *Europhysics Letters*, **138(2)**, 24002 (2022).

[27] Acosta, D. & Li, W., *Nuclear Instruments and Methods in Physics Research Section A: Accelerators, Spectrometers, Detectors and Associated Equipment*. **1027**, 166334 (2022).

[28] Acosta, D., Barberis, E., Hurley, N., Li, W., Colin, O. M., Wang, Y., ... & Zuo, X. *Journal of Instrumentation*, **18(09)**, P09025 (2023).

[29] B. Holzer, Presentation at "*ep/eA* versus *pp/pA/AA* synergy workshop", CERN 29 February - 01 March 2024; https://indico.cern.ch/event/1367865/

[30] U. Kaya, B. Ketenoglu and S. Sultansoy, in preparation.